\documentclass{jfm}
\usepackage{graphicx}
\usepackage{epstopdf, epsfig}
\usepackage{siunitx,amsmath}
\usepackage{amssymb,commath}
\usepackage{xcolor}
\usepackage{hyperref}
\usepackage{subfigure}

\newcommand{\comment}[1]{}

\shorttitle{Hydrogel sphere impact on granular media}
\shortauthor{Xiaoyan Ye and Devaraj van der Meer}

\title{Hydrogel sphere impact cratering, spreading and bouncing on granular media}

\author{Xiaoyan Ye\aff{1}\corresp{\email{yexy@lzu.edu.cn}} and Devaraj van der Meer\aff{2}}

\affiliation{\aff{1}Key Laboratory of Mechanics on Disaster and Environment in Western China attached to the Ministry of Education of China, and Department of Mechanics and Engineering Science, School of Civil Engineering and Mechanics, Lanzhou University, Lanzhou, Gansu 730000, PR China
\aff{2}Physics of Fluids Group, MESA+ Institute for Nanotechnology, and J.M. Burgers Center for Fluid Dynamics,University of Twente, P.O. Box 217, 7500 AE Enschede, The Netherlands}

\begin{document}

\maketitle

\begin{abstract}
The impact of a hydrogel sphere onto a granular target results in both the deformation of the sphere and the formation of a prominent topographic feature known as impact crater on the granular surface. We investigate the crater formation and scaling, together with the spreading diameter and post-impact dynamics of the spheres by performing a series of experiments, varying the Young's modulus  \(Y\) and impact speed \(U_{0}\) of the hydrogel spheres, and the packing fraction and grain size of the granular target.
We determine how the crater diameter and depth depend on \(Y\) and find the data to be consistent with those from earlier experiments using droplets and hard spheres. Most specifically, we find that the crater diameter data are consistent with a power law, where the power exponent changes more sharply when \(Y\) becomes less than $200$ Pa.  
Next, we introduce an estimate for the portion of the impact kinetic energy that is stored in elastic energy during impact, and thus correct the energy that remains available for crater formation. Subsequently, we determine the deformation of the hydrogel sphere and find that the normalized spreading diameter data are well collapsed introducing an equivalent velocity from an energy balance of the the initial kinetic energy against surface and elastic energy. 
Finally, we observe that under certain intermediate values for the Young's modulus and impact velocities, the particles rebound from the impact crater. We determine the phase diagram and explain our findings from a comparison of the elastocapillary spreading time and the impact duration.
\end{abstract}


\section{Introduction}
\label{sec:introduction}

Impact phenomena are ubiquitous in both nature and industry, and their scale ranges from the small dimensions of inkjet printing and raindrop dynamics to that of the impact of a large asteroid on a planet. Droplet impact on solid surfaces has been studied extensively ever since the pioneering work of Worthington~\citep{Worthington1908}. During impact, droplets spread over the solid surface, recede, splash, or rebound, depending on the properties of the intruders, substrate, and initial conditions, such as the impact velocity and impact angle \citep{Clanet2004,yarin2006,Josserand2016}. 

The observed phenomena become very different when the roles of substrate and impactor are reversed, i.e., when the substrate is deformable and the impactor is a solid object, such as during the impact of a sphere onto a granular bed \citep{Ruiz-Suarez2013,Katsuragi2016,Devaraj2017}. This process leads to the formation of an impact crater, and in 1993 already, \cite{Holsapple1993} pointed out the analogy between the morphology and dynamic processes of high-energy impacts of asteroids and other solar system bodies onto planets and laboratory-scale, low-energy impacts on granular materials. With the development of high-speed cameras, research on these low-energy impacts has thrived. Consequently, the literature on impact cratering in sand and its subsequent dynamics is quite extensive, ranging from the formation of the crater \citep{Uehara2003} and the study of its characteristics~\citep{Walsh2003,Zheng2004,Pacheco-Vazquez2019} to the force exerted on the impactor~\citep{Lohse2004,Katsuragi2007,Goldman2008} and the study of the splash that is created upon impact, also known as the ejecta~\citep{Marston2010,Pacheco-Vazquez2019}. 

Arguably as relevant as the above impact examples, is the interaction of a liquid droplet with a granular medium, such as the impact of a raindrop onto soil. Here, both impactor and substrate are deformable, and may even mix during the impact process. In the literature, the details of the crater formation \citep{Zhang2015,Jong2017,Liu2018}, liquid penetration and residue morphology~\citep{Delon2011,Katsuragi2011,Zhao2015} and maximum spreading diameter~\citep{Zhao2015a,Zhao2017} of the droplets have been investigated. 

It has become clear that craters formed by the impact of liquid and solid intruders are different. First of all, in a direct comparison between similar-sized droplets and solid spheres at similar impact energies, it was found that impact craters from liquid intruders are systematically wider and shallower from those created by the impact of a solid sphere \citep{Jong2020}. In this context, also the results of \cite{Pacheco2011} should be mentioned, who observed a similar behaviour for disintegrating granular impactors. Moreover, the diameter of craters formed by various types of intruders, were found to obey different scaling when fitted to a power law in the impact kinetic energy: For solid intruders the exponent is observed to be between \(0.2\)~\citep{Jong2020} and \(0.24\)~\citep{Uehara2003}, whereas the exponent is significantly lower for droplets, namely \(\sim 0.17-0.18\) \citep{Zhao2015a,Nefzaoui2012}.

This raises the question what would happen for an intruder with properties located in between a solid and a liquid, e.g., a highly deformable, elastic sphere. \cite{Matsuda2019} performed impact experiments onto granular media with packing fraction of \(0.618\) using elastic hydrogel spheres with a Young's modulus of several kPa, which indicate that the crater diameter obeys a power law in the impact kinetic energy with exponent \(0.223\) when the maximum indentation is smaller than 30\% of the sphere's diameter, and \(0.205\) for larger maximum indentations, thus indicating a value that, while it is still consistent with solid sphere impact, lies in between those for solid and liquid intruders mentioned above.

In the present study, we aim to extend these measurements to elastic hydrogel spheres with a broader range covering almost three orders of magnitude of Young's moduli, and directly compare our findings with those of liquid droplet and solid sphere impacts. We use granular substrates with two different size distributions and varying packing fractions for which we report both diameter and depth of the craters. Next, we measure the deformation behavior of the hydrogel spheres and introduce a similar energy decomposition as used in \citep{Zhao2015} to estimate the portions of the impact kinetic energy invested in sphere deformation and crater formation respectively. Moreover, we show that using this energy decomposition the maximum deformation data can be collapsed by means of the introduction  of an elasto-capillary Mach number, in a similar manner as was done for hydrogel spheres impacting on a solid substrate \citep{Tanaka2003,Tanaka2005,Arora2018}. Finally, we investigate the conditions under which the spheres rebound, and analyse it in terms of the spreading and impact time scales.

The paper is structured as follows. In the next Section~\ref{sec:expmethods} we discuss the experimental setup, the production of the hydrogel spheres and the measurement of the Young's modulus. In Section~\ref{sec:crater} we subsequently turn to the measurement of the crater characteristics and the comparison with liquid droplet and solid sphere impact. Here we also discuss the decomposition of the impact kinetic energy \(E_k\) into portions that are used for the deformation of the sphere (\(E_d\)) and the formation of the crater (\(E_s\)), respectively. We continue with the discussion of the maximum spreading diameter and spreading time of the sphere in Section~\ref{sec:spreading}, after which we turn to the bouncing phenomenon in  Section~\ref{sec:bouncing}. The paper is concluded in Section~\ref{sec:conclusion}.

\section{Experimental methods}
\label{sec:expmethods}
    
\begin{figure}
 \centerline{\includegraphics[width=\textwidth]{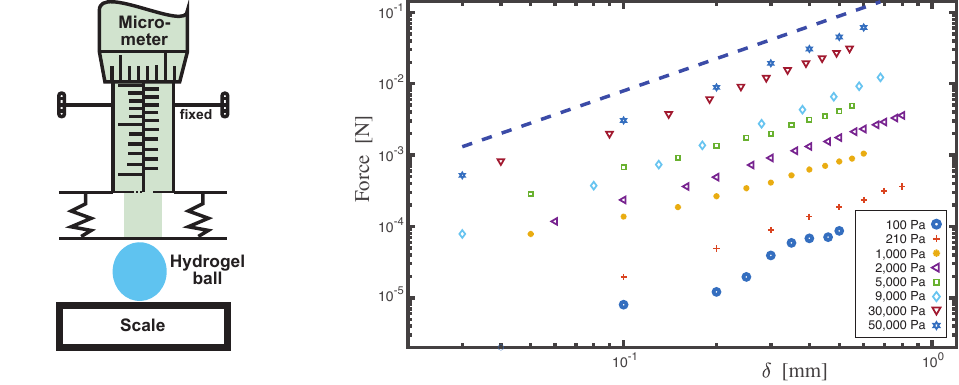}}
 \caption{ (\textit{a}) Sketch of the homemade Young's modulus test setup which is composed of a scales and a plate attached to a micrometer. The hydrogel balls are placed between the scales and the plate. The indentation \(\delta\) and the corresponding force \(F\) are measured while adjusting the micrometer. (\textit{b}) Force \(F\) and indentation \(\delta\) are plotted in a doubly logarithmic plot. The dashed line represents the power-law scaling with exponent \(3/2\) expected based on Hertz' theory.
}
 \label{fig:Youngs_Modulus}
 \end{figure}

In the experiments we investigate the impact of deformable hydrogel spheres onto granular materials by varying the impact velocity \(U_{0}\) and the stiffness of the spheres, as described by their Young's modulus \(Y\). We use a series of acrylamide gels that are prepared by the copolymerization of acrylamide (40\% stock solution) as monomer and methylenebisacrylamide (2\% stock solution) as comonomer. The amount of each reagent used for preparing the acrylamide gels is shown in Table \ref{tab:kd}. Sodium persulfate (\(0.93\) g/L) and tetramethylenediamine (\(0.6\) g/L) were added to initiate and accelerate the radical polymerization. The monomer and comonomer solutions were mixed before adding the initiators. Subsequently,  a volume of \(33.6\) \(\mu\)l of this solution (corresponding to a hydrogel sphere with a diameter \(D_0\) of approximately \(4.0\) mm) were transferred into an Eppendorf tube that was filled with polyoil of density \(1.006\) g/ml at \(25\) \(^\circ\)C to match that of the solution and allowed to polymerize. The Young's modulus \(Y\) of the hydrogel sphere was tuned by varying the concentrations of the monomer and comonomer and measured using a home-made device composed of a scales and a micrometer, as sketched in Fig.~\ref{fig:Youngs_Modulus}(a). The hydrogel sphere is placed between the scales and a rigid plate connected to a micrometer that was mounted on a stand. By adjusting the micrometer and neglecting the very small displacement of the scales, we obtain the indentation \(\delta\) of the sphere together with the corresponding force \(F\), as shown in Fig.~\ref{fig:Youngs_Modulus}(b). Now, the Young's modulus \(Y\) of the sphere may be determined using Hertz' theory for the deformation of an elastic sphere \citep{Matsuda2019}:
\begin{equation}\label{eq:Hertzlaw}
F(\delta)=\frac{4Y}{3\sqrt{2}(1-\nu^2)}D_{0}^{1/2}\delta^{3/2}\,,
\end{equation}
where the Poisson's ratio \(\nu=0.5\) was obtained invoking the incompressibility condition. We observe that the measurement data is well-fitted by the expected power law, with the exception of the data for the smallest values of \(Y\), which may be traced back to the fact that they suffer from very large deformation already at moderate forces. The measured Young moduli \(Y\) vary from approximately \(100\) to \(50,000\) Pa, as reported in Table \ref{tab:kd}.

\begin{table} \begin{center} 
\vspace{-0.5cm}\begin{tabular}{p{3cm} p{3cm} p{3cm} p{4cm}}
      Acrylamide [ml]  & Bisacrylamide [ml]    &   Water [ml] & $ Y \pm \Delta Y$ [kPa]\\
      (40\% stock solution)  & (2\% stock solution)    &   & \\
   \hline%\hline
       0.362 & 0.0465 & ~~4.5915 & $0.1 \pm 0.02$\\
       0.375 & 0.075  & ~~4.550   & $0.2 \pm 0.02$\\
       0.375 & 0.250  & ~~4.375  & $1.0 \pm 0.2$\\
       0.500   & 0.250   & ~~4.250   & $2.0 \pm 0.4 $\\
       0.500   & 0.5625 & ~~3.9375 & $5.0 \pm 0.4 $\\
       0.625 & 0.375  & ~~4.000    & $10.0 \pm 1.2$\\
       1.000 & 1.000  & ~~3.000    & $30.0 \pm 2.0$\\
       1.000   & 1.200    & ~~2.800    & $50.0 \pm 4.2$\\
        \hline
  \end{tabular}
  \caption{Young's modulus \(Y\) of hydrogel spheres obtained after polymerization of different acrylamide and bis-acrylamide concentrations. Here, \(\Delta Y\) represents the standard deviation in the modulus \(Y\) obtained from different batches of hydrogel spheres obtained with the same preparation method. Note that the variation of the modulus within a batch of particles is significantly smaller than  \(\Delta Y\) and therefore more precise values may be found within this study. Also note that not all batches reported in this table have been used for experiments reported in this manuscript.}
  \label{tab:kd}
\end{center}
\end{table}

For the impact experiments, these hydrogel spheres are released from a tiny needle connected to a low-pressure pump. They are impacted onto a granular bed with an impact speed \(U_0\) varying from \(0.5\) to \(4.5\) m/s, which is computed from the images obtained with a high-speed camera at a frame rate of 10,000 fps. 
White hydrophobic silane-coated soda lime beads are used as granular target, of which the density is equal to \(\rho_{g} = 2.5\cdot10^{3}\) kg/m\(^{3}\). We use two different polydisperse batches of grains, with a mean size of \(200\) \(\mu\)m and \(114\) \(\mu\)m, respectively. In this work, we mainly focus on results obtained with the \(200\) \(\mu\)m grains. The beads are dried in an oven at around \(105\) \(^\circ\)C for at least half an hour and subsequently cooled down before the experiment. The packing fraction of the target is tuned in the range of \(0.55 - 0.62\) by air fluidization and subsequent tapping, as shown in Fig.~\ref{fig:snapshots}. The packing fraction is determined by measuring the surface profile with a laser profilometer just before each impact experiment. Two high-speed cameras are employed to simultaneously record the deformation of the hydrogel balls and the crater profile, determined with 
an in-house developed high-speed laser profilometry setup discussed extensively in~\citep{Zhao2015}, from which the deformation of the granular bed can be deduced. The dynamic crater profile \(z(r,t)\) is captured using this high-speed laser profilometer under the assumption of axisymmetry, and the spreading diameter of the hydrogel sphere is detected simultaneously by monitoring the time evolution of the contour of the sphere in the images acquired by the other camera.
  
 \begin{figure}
  \centerline{\includegraphics[width=0.6\textwidth]{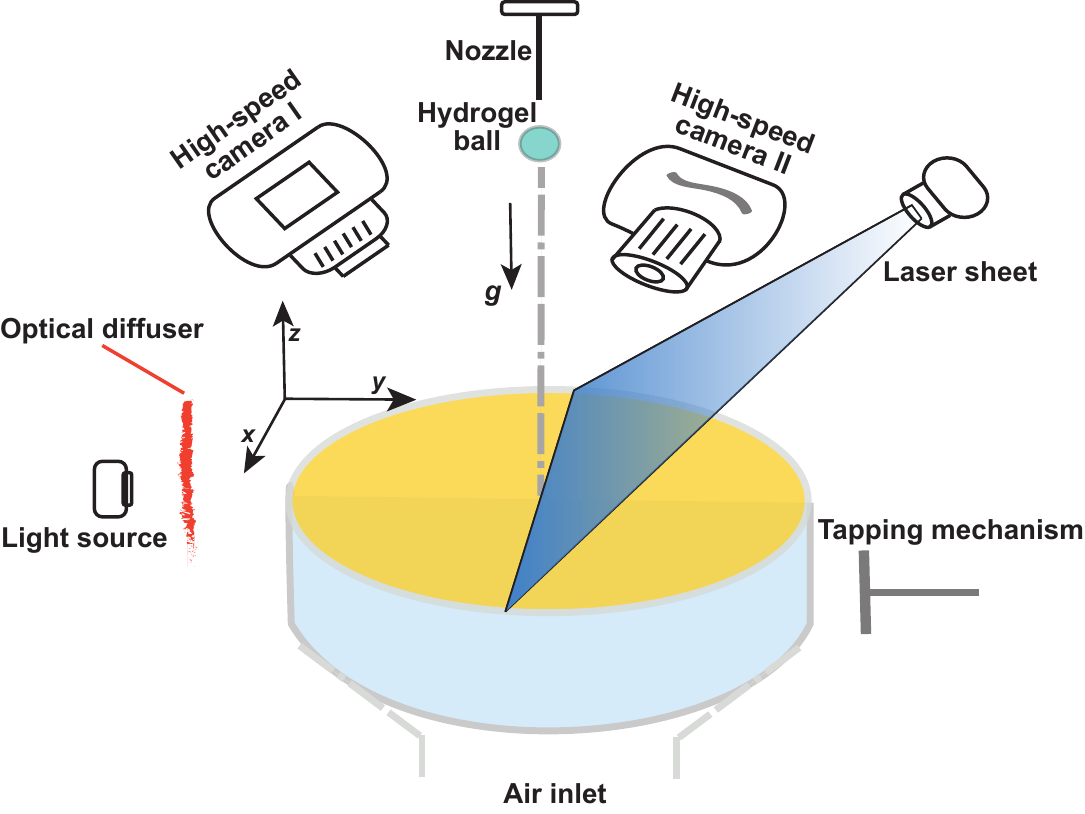}}
    \caption{Sketch of the experimental setup. Before impact a granular target is prepared in a given packing fraction by fluidization with dry air and subsequent tapping. The impact of the hydrogel sphere and its deformation is subsequently imaged with one high-speed camera, whereas simultaneously, the deformation \(z(x,y,t)\) of the granular bed is measured using two laser sheet imaged by a second high-speed camera. A two-dimensional scan of the surface is made before and after the experiment using the same laser sheets and a translation stage (not depicted) providing the three-dimensional crater profile.}
   \label{fig:setup}
  \end{figure}

\section{Target deformation: Crater characteristics}
\label{sec:crater}

\begin{figure}
\centerline{\includegraphics[width=\textwidth]{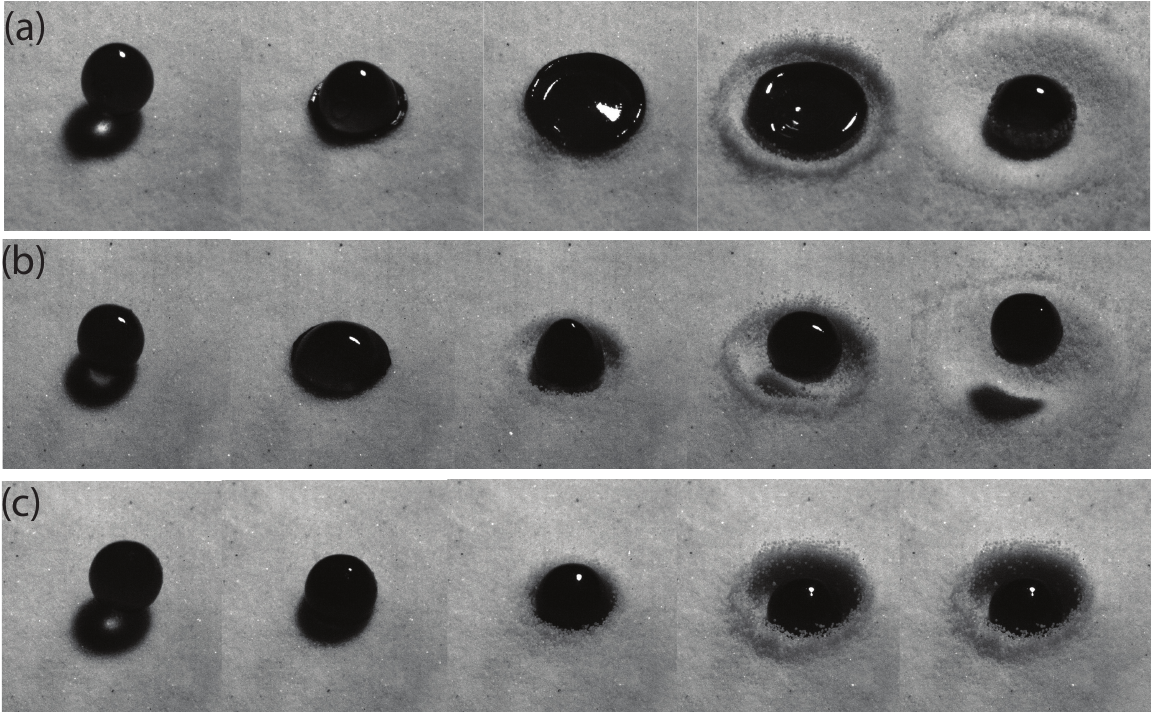}}
\caption{Snapshots from the impact of hydrogel spheres onto a sand bed with an average grain diameter of $114$ $\mu$m, for three different values of the Young's modulus of the sphere, namely (a)$100$ Pa, (b) $5,000$ Pa, and (c) $30,000$ Pa, each for the same impact velocity of \(U_0 = 1.5\) m/s. 
Each row represents five consecutive snapshots of an impact, where the leftmost column illustrates the moment of initial contact between the hydrogel sphere and the granular target,  $t=t^*$. The subsequent snapshots correspond to $t=t^*+10$ ms, $t=t^*+30$ ms, $t=t^*+70$ ms, and $t=t^*+270$ ms.}
\label{fig:snapshots}
\end{figure}

Three typical examples of the impact of hydrogel spheres of varying softness onto the granular target are shown in Fig.~\ref{fig:snapshots}, where the Young's modulus \(Y\) increases from top to bottom. In all three cases, the sphere is observed to expand radially to its maximum diameter \(d_{max}\) and then recedes; while the interaction force and post-motion become more complex. Fig.~\ref{fig:snapshots}a illustrates how the impact process of a very soft sphere (\(Y=100\) Pa and \(U_{0}=1.5\) m/s) is similar to that of droplets~\citep{Zhao2017}, except for the fact that the hydrogel sphere is of course not able to penetrate into the sand bed and becomes covered with sand particles that stick to its surface, similar to what happens during water droplet impact on hydrophobic sand. A pronounced lamella develops and spreads outward while the bulk of the sphere continues to move downward as a truncated sphere. Meanwhile, the lamella continues to expand, developing a thicker rim at the edge until it reaches its maximal radius. A retraction phase follows where the hydrogel recovers its shape moving up in the vertical direction and eventually comes at rest under the influence of gravity. Note that the formation of a rim indicates that interfacial tension may be important for the dynamics of the sphere at this very small value of the Young's modulus, a fact that we will address in greater detail when discussing the deformation of the spheres in the next Section~\ref{sec:spreading}. 

Snapshots of the motion of a hydrogel sphere of moderate stiffness (\(Y=5,000\) Pa and \(U_{0}=1.5\) m/s) are presented in Fig.~\ref{fig:snapshots}b. 
Shortly after impact, the shape of the visible upper part of the soft ball is similar to that of a half ellipsoid. Even when the spreading radius reaches its maximum value, the central thickness of the sphere is still significantly larger than its rim. Note that this behavior is qualitatively different from that observed in Fig.~\ref{fig:snapshots}a where the sphere spreads in its entirety before retracting. A retraction phase follows where the hydrogel elongates in the vertical direction and eventually rebounds and completely loses contact with the granular bed, as can be appreciated from the shadow cast by the sphere onto the granular bed in the last two snapshots. The bouncing phenomenon will be discussed in greater detail in Section~\ref{sec:bouncing}. 

Finally, for one of the stiffest spheres we investigated in this study (\(Y=30,000\) Pa and \(U_{0}=1.5\) m/s) the snapshots of the impact process are shown in Fig.~\ref{fig:snapshots}c. Upon contact, the hydrogel ball exhibits no significant spreading phase at this impact speed and the impact resembles that of a solid ball. At much smaller impact speeds, where the impact region becomes visible, it appears that the sphere deforms at most locally or even shows no obvious deformation. Note that for these stiffest spheres no bouncing is observed after the impact.

\subsection{The morphology of impact crater}
\label{subsec:cratermorphology}

Craters are formed on the surface of the granular bed after impact of the soft spheres, the diameter and depth of which is influenced by the Young's modulus of the spheres. The crater morphology is deep and narrow for stiffer balls, while it is shallow and wide for softer balls. In the next subsections we will address the question how the stiffness of the ball and (to lesser extent) how the packing fraction of the granular bed affect the crater morphology. More specifically, we will investigate whether the dynamics of solid and liquid intruders can be considered as the limiting cases of that of very stiff hydrogel spheres (\(Y\rightarrow\infty\)) and very soft ones (\(Y\rightarrow 0\)), respectively. 
 
\begin{figure}
  \centerline{\includegraphics[width=0.8\textwidth]{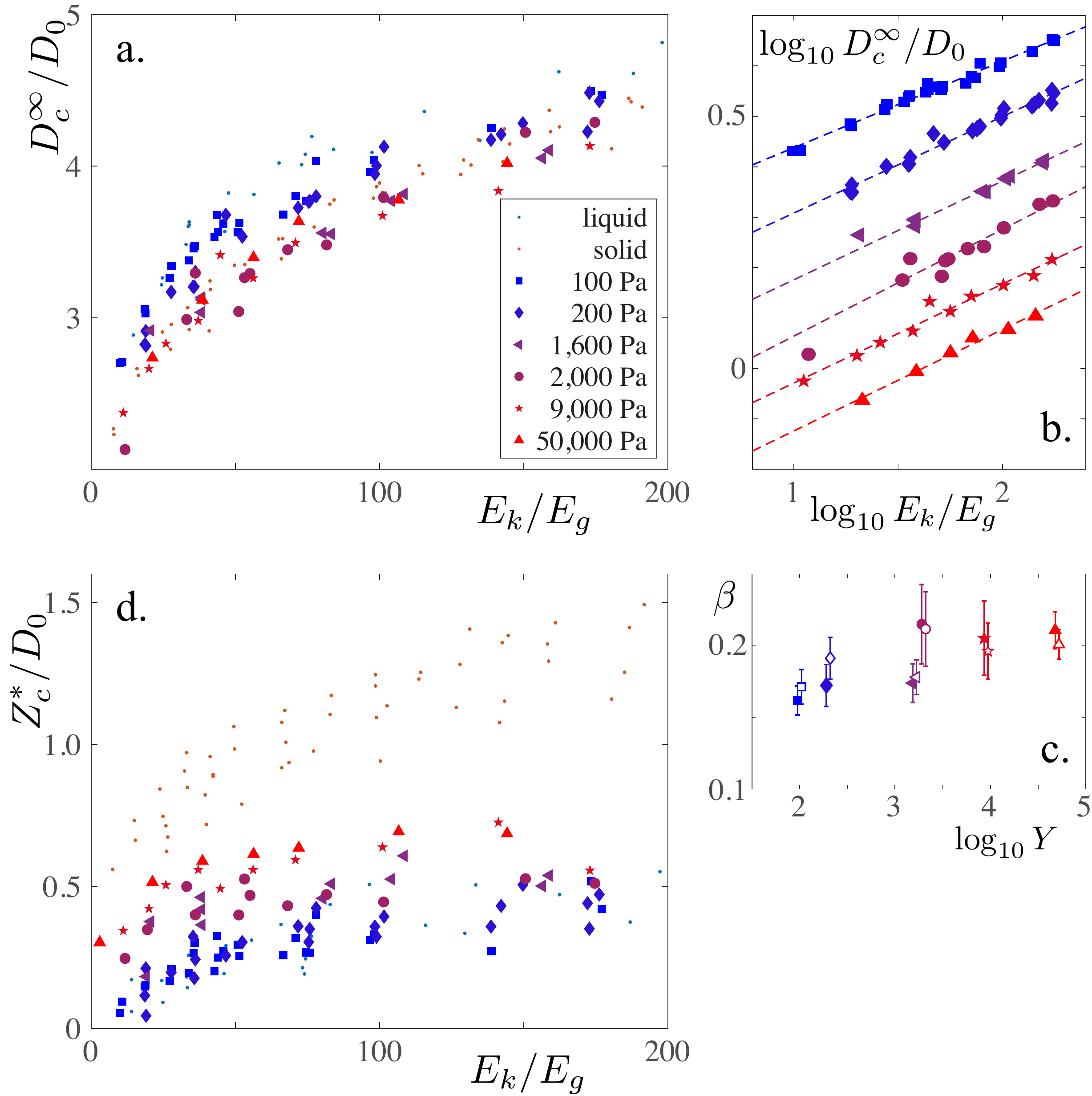}}
  \caption{(a) The dimensionless maximum crater diameter  \(D_{c}^\infty\) versus the non-dimensionalized kinetic energy \(E_{k}/E_{g}\) of the impact of hydrogel spheres with different Young moduli \(Y\) onto a granular bed with average grain diameter of 200 \(\mu\)m and a packing fraction in the range of \(0.585-0.62\). The data are compared to those obtained in \cite{Jong2020} for the impact of a solid ball (red dots) and a water droplet (blue dots) of similar size on a granular bed of similar grain size and packing fraction (see legend). (b) The same data as in a, but now plotted doubly logarithmically to verify behavior consistent with a power law. Note that the data have been shifted in the vertical direction to avoid overlapping of the data for different values of \(Y\). (c) Power-law exponents \(\beta\) (closed symbols) obtained for the data presented in b, as a function of the Young's modulus \(Y\) of the hydrogel spheres. The open symbols are power-law-exponent data obtained from corrected kinetic energy \(E_s/E_g\), where Young moduli have been slightly shifted for visualization purposes. (d) The dimensionless crater depth \(Z_{c}^\ast\) versus the non-dimensionalized kinetic energy \(E_{k}/E_{g}\) for the same data as presented in a, under the same conditions and again compared to the solid ball (red dots) and droplet (blue dots) data. Note that the legend of a is shared by the other figures.
}
\label{fig:DiamDepthvsEk}
\end{figure}
 
\subsection{Crater diameter \(D_{c}^\infty\)}
\label{subsec:craterdiameter} 

The crater diameter \(D_{c}^\infty\) is defined as the radial diameter of the top rim of the crater, taken at a moment in time after all of the dynamics associated to the hydrogel sphere impact has ceased. Fig.~\ref{fig:DiamDepthvsEk}a shows the dimensionless final crater diameter \(D_{c}^\infty/D_{0}\), i.e., rescaled with the sphere diameter \(D_0\), versus the dimensionless impact kinetic energy \(E_{k}/E_{g}\) for hydrogel spheres that compared with data obtained for solid balls and droplets under similar conditions in~\citep{Jong2020}. Here, \(E_{k}=(\pi/12)\rho_{i}D_{0}^3U_{0}^2\) is the kinetic energy just before the sphere comes in contact with the granular bed, and \(E_{g}=(\pi/6)\rho_{g}\phi_{0}gD_{0}^4\) is the gravitational potential energy associated with the motion of bed particles with a volume equivalent to that of the sphere is moved over a distance corresponding to the sphere diameter \(D_{0}\) vertically. Clearly, \(D_{c}^\infty/D_{0}\) increases with \(E_{k}/E_{g}\) for all Young moduli investigated. Moreover, for fixed \(E_{k}/E_{g}\), the crater diameter tends to increase with decreasing \(Y\). The hydrogel sphere data for \(D_{c}^\infty/D_{0}\) is bounded on the upper side by the data for the droplet and on the lower side (although somewhat less convincingly) by the solid intruder data. For the hydrogel sphere with the lowest Young modulus (\(Y=100Pa\), the diameter of the resulting crater is consistent with that of an impacting droplet when \(E_{k}/E_{g}<100\). When \(E_{k}/E_{g}>100\), the crater diameter of soft hydrogel spheres is smaller than that of droplets, possibly due to the sphere not spreading as strongly as the droplet, and certainly not splashing. With the increase of \(Y\), the data shifts towards and even significantly below that obtained for a rigid sphere. This observation appears surprising at first sight, but will be revisited after introducing the reduced impact energy \(E_s\) in Subsection~\ref{subsec:correctedenergy}.

When we plot the same data in a doubly logarithmic scale (Fig.~\ref{fig:DiamDepthvsEk}b), where data is shifted in the vertical direction for each value of \(Y\) for clarity, we observe that the data is consistent with a power law of the form \(D_{c}^\infty/D_0 \sim (E_{k}/E_{g})^{\beta}\). In (Fig.~\ref{fig:DiamDepthvsEk}c) the corresponding power-law exponents \(\beta\), obtained from a linear best fit to the data in Fig.~\ref{fig:DiamDepthvsEk}b, are plotted versus the logarithm of the Young's modulus \(Y\). Clearly, \(\beta\) increases with increasing \(Y\) from a value close to \(\beta=0.17\) for low Young moduli, which is consistent with that observed for droplet impacts, to values just above \(\beta=0.21\), which corresponds to power-law exponents of solid sphere impact on sand reported in the literature.
Finally, it is good to note that the exponent obtained for \(Y = 2,000\) Pa appears to be an outlier, when compared to the data obtained for the other Young moduli, which is consistent with the large error margin of the best fit.

\subsection{Crater depth \(Z_{c}^\ast\)}
\label{subsec:craterdepth} 

The maximum crater depth \(Z_{c}^\ast\) 

is obtained from the final crater profile by subtracting the height of the hydrogel sphere \(h\) from the centre of the crater profile \(Z(0,t^\infty)\) (in the case that the hydrogel sphere had not bounced out of the crater). This is consistent with the method with which the crater depth has been determined in~\citep{Jong2020} and in numerous earlier papers determining the crater depth for solid sphere impact. 

Note that the height \(h\) is not exactly equal to \(D_0\), but obtained by subtracting the indentation \(\delta_0\) induced by gravity from \(D_{0}\). 
The dimensionless crater depth \(Z_{c}^\ast/D_0\) is plotted against \(E_{k}/E_{g}\) in Fig.~\ref{fig:DiamDepthvsEk}d. Obviously, the crater is deeper for stiffer beads, indicating that \(Z_{c}^\ast\) increases with \(Y\), and its range varies between the impact depth of solid ball and droplet, as one would expect, with the data for small \(Y\) close to that of the droplet and increasing towards that of the solid sphere impact for larger \(Y\). However, it should not remain unmentioned that, whereas the impact depth for the softest spheres (\(Y = 100\) Pa) is basically coincident with that of the droplet, the data for the largest Young's modulus, \(Y = 50,000\) Pa, do not come close to that of the solid sphere impact. It could well be that even the slight deformation that the stiffest hydrogel sphere experiences adds to the friction that the sphere experiences inside the sand, thereby significantly decreasing its maximum penetration depth.

 \subsection{Correcting the impact kinetic energy for sphere deformation.}
\label{subsec:correctedenergy} 
 
\begin{figure}
  \centerline{\includegraphics[width =0.7\textwidth]{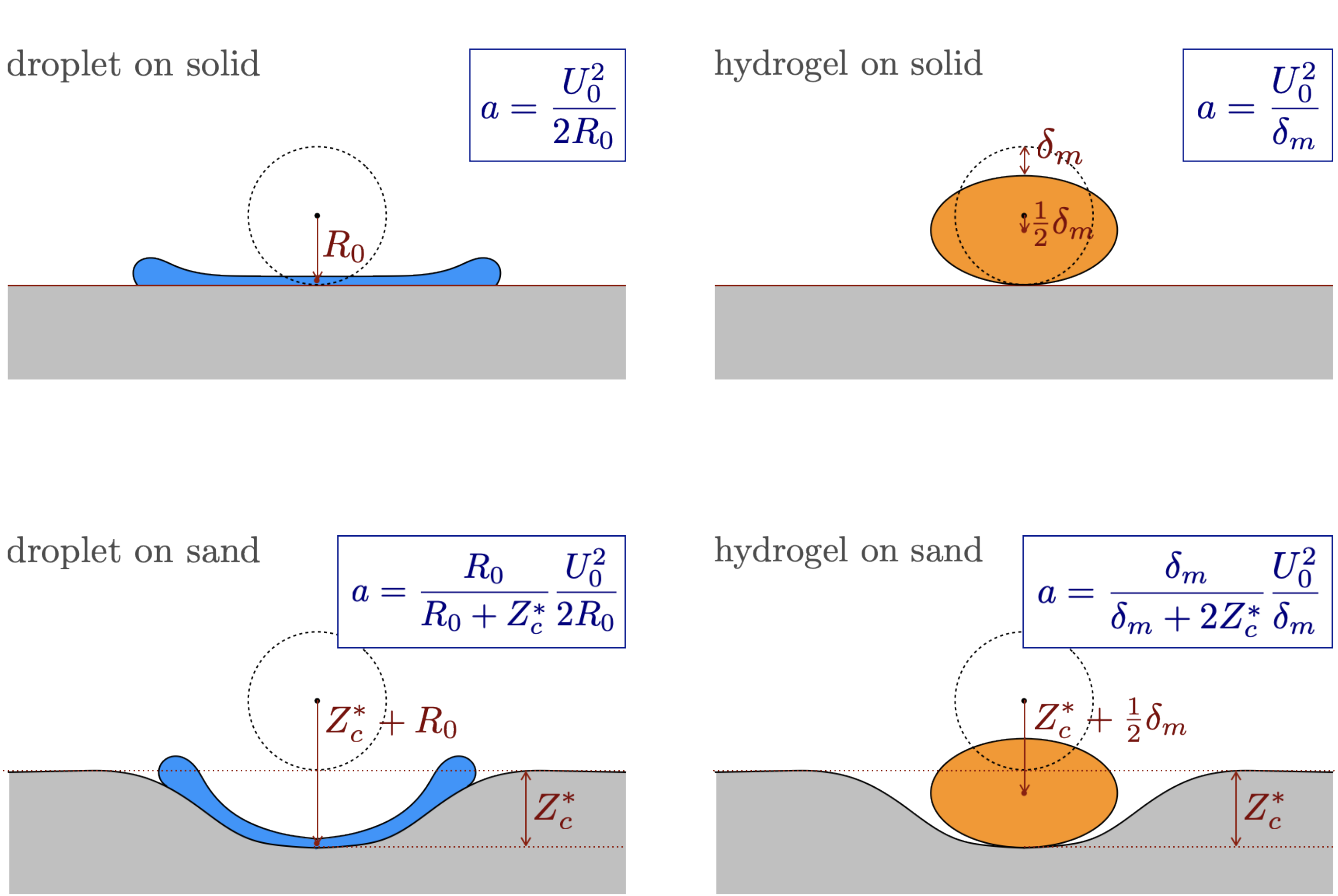}}
  \caption{The relevant average deceleration experienced by (a) a droplet impacting on a solid, (b) a hydrogel sphere impacting on a solid, (c) a droplet impacting on sand, and (d) a hydrogel sphere impacting on sand, respectively. Here, \(R_{0}\) is radius of the droplet, which is approximately equal to the deformation of droplet during impact onto a solid, \(\delta_{m}\) is the maximum indentation of the hydrogel sphere, such that \(\delta_{m}/2\) is the displacement of its centre of mass, and \(Z_{c}^\ast\) is the maximum crater depth reached during impact.
 }
\label{fig:decelerationschematic}
\end{figure} 
 
As argued in our previous work, during impact of a droplet part of the impact energy is used for the deformation of the droplet such that only the remaining part may be used for the formation of the crater. The same argument holds for the impact of a hydrogel sphere, where part of the impact energy is used for the elastic deformation of the sphere. Following the reasoning of \cite{Clanet2004} what happens when a droplet impacts onto a solid substrate is that it deforms as a result of a deceleration force that it experiences during impact. This force performs an amount of work on the droplet along the displacement of the center of mass, which in this case is approximately equal to the droplet radius \(R_{0} = D_0/2\), as depicted in Fig.~\ref{fig:decelerationschematic}a, assuming that the droplet spreads until it has a negligible thickness in its centre. Since at maximum spreading of the droplet no vertical momentum is left in the droplet, this amount of work must have transformed the impact kinetic  energy \(E_{k}\) into other forms of energy, in this case the surface energy of the droplet. In this way, the average force exerted by the solid substrate onto the droplet is equal to the impact kinetic energy divided  by the displacement, i.e., equal to \(E_{k}/R_{0}\), such that the average deceleration \(a\) experienced by the droplet may be written as \(a \approx U_{0}^2/(2R_{0})\). 
 
When the same droplet however impacts onto a granular bed, the force that deforms the droplet acts over a larger distance, namely the sum of the droplet radius and the maximum crater depth \(Z_{c}^\ast\), and as a consequence the average deceleration is smaller in magnitude, \(a\approx U_{0}^2/(2(R_{0}+Z_{c}^\ast))\) (Fig.~\ref{fig:decelerationschematic}c). This implies that with the same impact kinetic energy, the deceleration drops from \(a\approx U_{0}^2/(2R_{0})\) to \(a \approx U_{0}^2/(2(R_{0}+Z_{c}^\ast))\), i.e., by a factor \(R_0/(R_{0}+Z_{c}^\ast)\), as if the droplet had impacted a solid substrate with a lower impact velocity, corresponding to a kinetic energy \(E_d = R_0/(R_{0}+Z_{c}^\ast)\,E_k\). Stated in a slightly different manner, one could say that a portion \(E_d = R_0/(R_{0}+Z_{c}^\ast)\,E_k\) of the impact kinetic energy \(E_k\) is used for droplet deformation, and the remaining portion \(E_s = E_k - E_d = Z_{c}^\ast/(R_{0}+Z_{c}^\ast)\,E_k\) is used for the formation of the crater \citep{Zhao2015}.

We now use the same argument in the case of a hydrogel sphere. When the sphere impacts onto a solid substrate and vertically deforms over a maximum length \(\delta_m\), the deceleration working can be found similarly as for the droplet case: Assuming that the point of maximum deformation  is reached when there is kinetic energy left in the vertical direction, one needs to divide the kinetic energy of the impactor by the displacement of the centre of mass, that is, by \(\delta_m/2\) and divide by the impactor mass to obtain the deceleration \(a\) (Fig.~\ref{fig:decelerationschematic}b), i.e.,
\(a = U_0^2/\delta_m\).

Clearly, when the solid substrate is replaced by a sand bed, the displacement of the centre of mass of the sphere is now equal to \(\delta_m/2+Z_{c}^\ast\) and consequently the magnitude of the deceleration is decreased to
\begin{equation}\label{eq:deceleration}
a = \frac{U_0^2}{\delta_m+2Z_c^\ast} = \frac{\delta_m}{\delta_m+2Z_c^\ast}  \,\frac{U_0^2}{\delta_m}\,,
\end{equation}
that is, again, the deformation of the hydrogel sphere impacting onto the granular bed is equal to that it would have obtained when the sphere would have impacted a solid substrate at a lower kinetic energy, obtained by multiplying the impact kinetic energy by a factor \(\delta_m/(\delta_m+2Z_c^\ast)\). This leads to a distribution of the impact kinetic energy \(E_k\) in a part \(E_d\) that is being used for the deformation of the hydrogel sphere and another part \(E_s\) (\(= E_k - E_d\)) that is invested in the formation of the the crater, respectively given by 
\begin{equation}\label{eq:energydistr}
E_d = \frac{\delta_m}{\delta_m+2Z_c^\ast}E_k \qquad\textrm{and}\qquad E_s = \frac{2Z_c^\ast}{\delta_m+2Z_c^\ast}E_k\,.
\end{equation}
When we assume that the hydrogel sphere maintains an ellipsoid shape during impact, we may calculate its maximum indentation \(\delta_{m}\) based on the incompressibility condition, that is \(\delta_{m} \approx D_{0}-{D_{0}^3}/{d_{max}^2}\), in which \(d_{max}\) is the maximum deformation diameter of the hydrogel sphere, which will be discussed extensively in the next Section~\ref{sec:spreading}. 

\begin{figure}
  \centerline{\includegraphics[width =\textwidth]{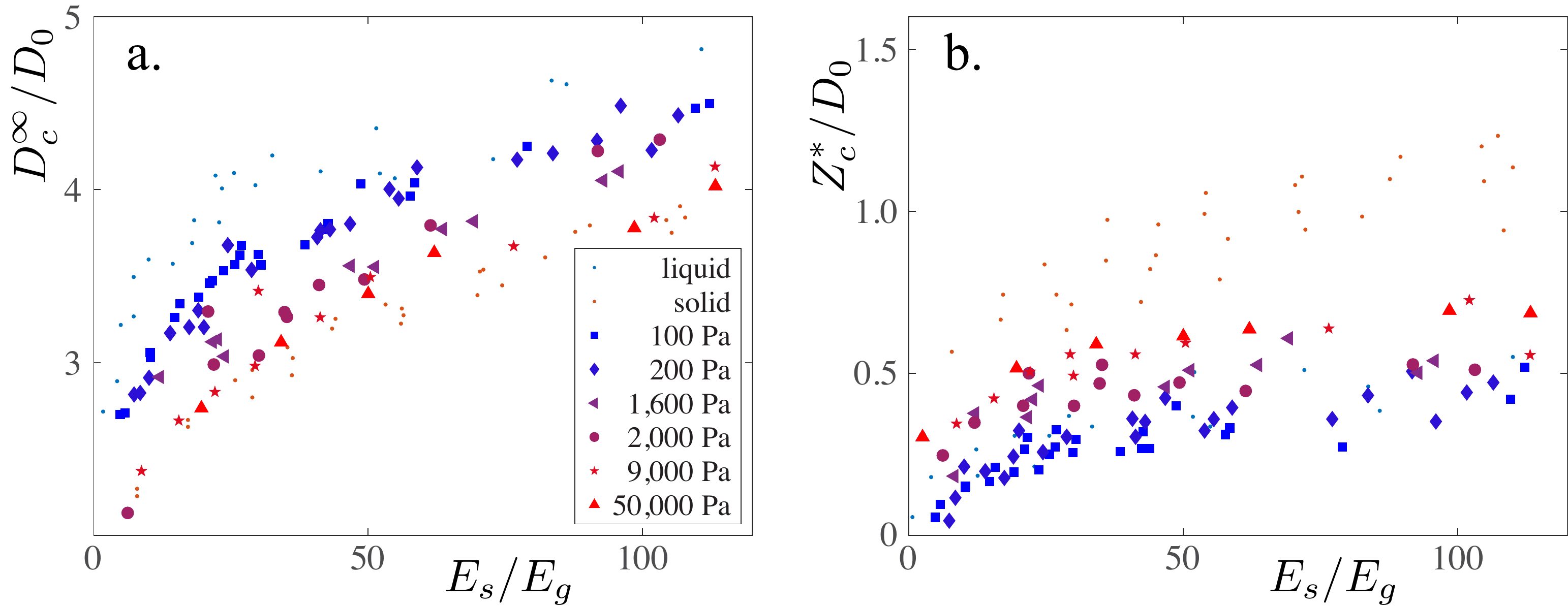}}
  \caption{(a) Rescaled crater diameter \(D_{c}^\infty/D_0\) and (b) rescaled crater depth \(Z_{c}^\ast\) versus the rescaled portion of the impact kinetic energy going into sand deformation \(E_{s}/E_{g}\). The data in these plots are the same as presented in Fig.~\ref{fig:DiamDepthvsEk}, where the small red and blue dots respectively represent the solid sphere and droplet impact published in \cite{Jong2020} and the large symbols the current hydrogel sphere data.
}
\label{fig:DiamDepthvsEs}
\end{figure}

With measured \(d_{max}\), \(Z_{C}^\ast\) and \(E_{k}\), the energy portion consumed by the substrate \(E_{s}\) can be computed for all experiments. In Fig.~\ref{fig:DiamDepthvsEs}a we plot the rescaled crater diameter \(D_{c}^\infty/D_{0}\) against \(E_{s}/E_{g}\). We find that the gap between the data points corresponding to the liquid and solid intruder is widened, with the data of the hydrogel spheres in between arranged strictly according to their Young moduli \(Y\). Most importantly, where in Fig.~\ref{fig:DiamDepthvsEk}a there was significant overlap between the solid sphere data and those of the stiffest hydrogel spheres, this overlap has disappeared after correcting the impact kinetic energy: Now all hydrogel data convincingly lies between both the droplet-impact data and the solid-sphere-impact data, confirming that (at least from the crater-diameter perspective) solid sphere and droplet impact can indeed be interpreted as limiting hydrogel sphere impact for the stiffest (\(Y\to\infty\)) and the softest (\(Y\to 0\)) case: Clearly, the crater diameter obtained by the impact of the softest hydrogel sphere with Young's modulus of the order of a hundred Pa is close to that of the droplet, whereas the case with \(Y\) of the order of several 10,000 Pa is close to that of a solid sphere in the low-energy impact area, and the crater diameter is increased in the high-energy range due to a more significant deformation of hydrogel sphere. The above observations build trust in the correctness of the energy decomposition scenario introduced in this Subsection, such that we will continue to use it in the rest of this study.

Subsequently, we also replotted the crater diameter data versus the corrected energy on a doubly logarithmic scale, verified that again the data is consistent with a power law (not shown) and determined the exponents, which are added as the open symbols in Fig.~\ref{fig:DiamDepthvsEk}c and found to lie very close to those determined from the uncorrected impact kinetic energy (the closed symbols in the same plot). The only difference is that the exponents appear to be slightly higher for small \(Y\) and slightly smaller for large \(Y\), but no clear distinction can be made within the measurement error.   

Finally, in Fig.~\ref{fig:DiamDepthvsEk}b we plot the rescaled crater depth as a function of the corrected and rescaled impact kinetic energy \(E_s/E_g\). Here we clearly observe a large discrepancy between the solid-sphere impact data and that of even the stiffest hydrogel spheres. This we tentatively attribute to the different friction that may be experienced by the hydrogel spheres inside the sand bed, combined with the fact that, even for the stiffest hydrogel spheres, there is a slight deformation of the sphere upon impact that may further increase the friction that they experience. The fact that also the softest hydrogel spheres appear to undergo increased friction moving into the bed supports this viewpoint. This increased friction may be connected to the observed tendency of the sand grains to slightly stick to the surface of the hydrogel spheres.

\section{Impactor deformation: Maximum spreading diameter}
\label{sec:spreading}

\begin{figure}
  \centerline{\includegraphics[width=\textwidth]{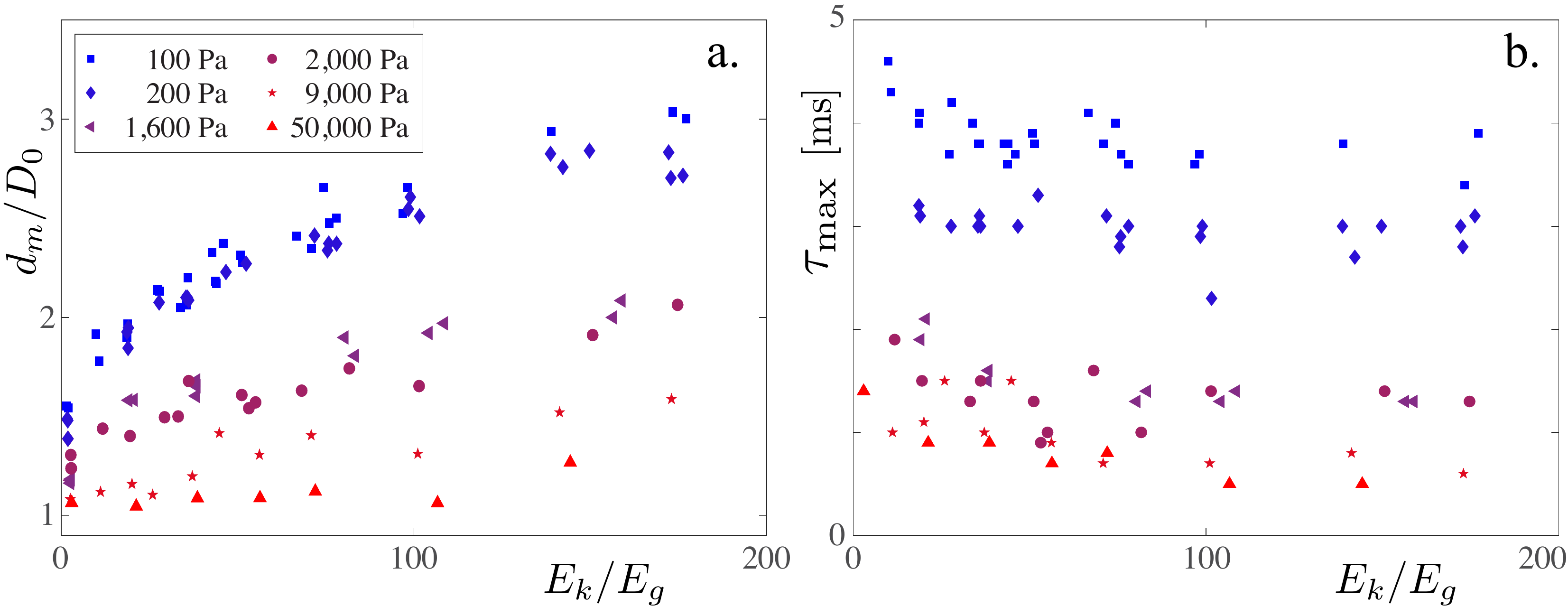}}
  \caption{(a) Dimensionless maximum spreading diameter \(d_{m}/D_{0}\) of the hydrogel spheres, and (b) spreading time \(\tau_\textrm{max}\), versus dimensionless impact kinetic energy \(E_{k}/E_{g}\) for the impact of hydrogel spheres with different Young moduli \(Y\) onto a granular bed with average grain diameter of 200 \(\mu\)m and a packing fraction in the range of \(0.585-0.62\), i.e., for the same parameters as presented in Figs.~\ref{fig:DiamDepthvsEk} and~\ref{fig:DiamDepthvsEs}.}
 \label{fig:SpreadingDiameterVsEk}
\end{figure}

Deformation occurs simultaneously for both the granular bed and the hydrogel sphere during impact. For the hydrogel spheres there are three deformation regimes within the concerned velocity range: the concave pancake regime, the quasi-ellipsoid regime and the local deformation or Hertz regime \citep{Tanaka2003}. The three regimes are clearly visible in Fig.~\ref{fig:snapshots}, where the concave pancake-like deformation appears in the third and fourth snapshot of series a, the quasi-ellipsoid regime is visible in the second snapshot of series b, and finally the local deformation may be inferred in series c from the fact that there is no apparent change in the horizontal diameter of the sphere during the entire impact.

To trace the morphological evolution of the hydrogel spheres we adopted the method presented in~\citep{Zhao2019}. Due to the perspective of the camera and lighting, the sphere appears as a dark ellipse in the experimental images. To detect the deformation of the hydrogel sphere, cross-correlations are computed between dark ellipse templates and the (background subtracted) image, and the template size corresponding to the maximum correlation provides the spreading diameter \(d\) in the direction perpendicular to the impact. From the entire time series we subsequently determine the maximum spreading diameter \(d_m\).

In Fig.~\ref{fig:SpreadingDiameterVsEk}a we plot the maximum spreading diameter \(d_{m}\) scaled by the sphere diameter $D_0$ as a function of the dimensionless impact energy $E_k/E_g$, for the same Young moduli \(Y\) and impact velocities \(U_0\) as for which we determined the crater characteristics plotted in Figs.~\ref{fig:DiamDepthvsEk} and~\ref{fig:DiamDepthvsEs}. Two things are apparent in this Figure, namely that the maximum spreading diameter increases significantly with decreasing \(Y\) and, secondly, that with increasing impact kinetic energy \(d_m\) increases as well. We even observe that for the stiffest sphere (\(Y = 50,000\) Pa), where for low \(E_k\) there is no observable horizontal deformation, the impact is able to generate global deformation for the largest impact energy. For the smallest Young modulus (\(Y = 100\) Pa) the deformation reaches up to 200\% of its original size, comparable to the deformations experienced by impacting liquid droplets.

\begin{figure}
\centerline{\includegraphics[width=\textwidth]{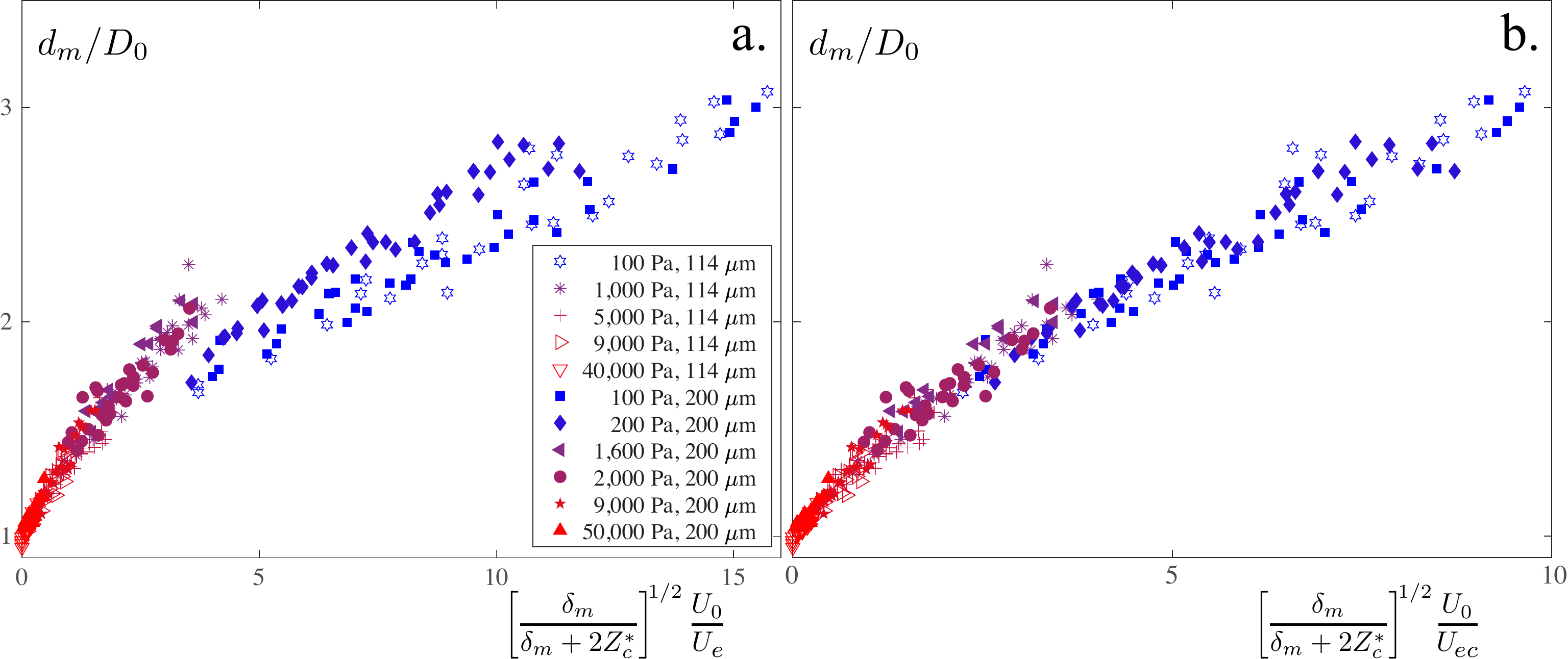}}
\caption{The dimensionless maximum spreading diameter \(d_{m}/D_{0}\) of the impacting hydrogel spheres is plotted against (a) the modified elastic Mach number \((\delta_{m}/(\delta_{m} + 2Z_{C}^\ast))^{1/2}(U_0/U_e)\) and (b) the modified elastocapillary Mach number \((\delta_{m}/(\delta_{m} + 2Z_{C}^\ast))^{1/2}(U_0/U_{ec})\) as explained in the text. Note that the data set of Fig.~\ref{fig:SpreadingDiameterVsEk} (i.e., for 
average grain diameter of \(200\) \(\mu\)m and a packing fraction in the range of \(0.585-0.62\)) has now been enlarged to also include smaller average grain diameter (\(114\) \(\mu\)m) and smaller packing fractions (in the range \(0.55-0.585\)).  
}
\label{fig:SpreadingDiameterVsMa}
\end{figure}

Clearly, there is little additional order to be observed in Fig.~\ref{fig:SpreadingDiameterVsEk}a, which suggests that $E_k/E_g$ is not the most appropriate parameter to describe the deformation of the spheres during impact. The first step would be use the portion of the impact kinetic energy that is available for the sphere, i.e., \(E_d = E_k- E_s = (\delta_{m}/(\delta_{m} + 2 Z_c^\ast))E_k\), instead of \(E_k\), but plotting \(d_{m}/D_{0}\) versus \(E_d/E_g\) does not bring much. This is easily understood, since clearly the elastic deformation should be a main ingredient of a more coherent description, for which we turn to the literature. \cite{Tanaka2003} reported on the impact of a cross-linked gel with \(Y\) around \(2,000\) Pa onto a non-sticking, rigid substrate. They showed that the spreading dynamics can be rationalised from a balance between inertia and bulk elasticity, \(E_k = E_e\), where \(E_k\) is the kinetic energy just before impact and \(E_e\) is the elastic energy stored in the sphere at its maximum deformation, namely at the point where the kinetic energy can be assumed to be zero. Starting from the analysis described in \cite{Arora2018} we express the stored bulk elastic energy as a function of the maximum deformation  \(\lambda_m=d_m/D_{0}\) as \(E_{e} \approx \tfrac{1}{12}\pi D_{0}^3 G(2\lambda_m^2+\lambda_m^{-4}-3)\), where \(G\) is the shear modulus of the sphere, which is related to Young's modulus \(Y\) as \(Y=2G(1+\nu)\approx3G\) where, again, we took the Poisson ratio \(\nu \approx 0.5\). When the maximum spreading diameter is very large (\(\lambda_m \gg 1\)), the first term dominates the elastic energy, which may than be approximated as  \(E_{e} \approx \tfrac{1}{6}\pi D_{0}^3 G \lambda_m^2\). 
Translated to the current situation, we need to equate \(E_e\) with the portion \(E_d\) of the impact kinetic energy that is available for the sphere deformation, which leads to \(\lambda_m^2 \approx \tfrac{1}{2}(\delta_{m}/(\delta_{m} + 2 Z_c^\ast)) \rho U_0^2/G\), or
\begin{equation}\label{eq:lambdam}
\lambda_m = \frac{d_m}{D_0} \approx \frac{1}{\sqrt{2}}\sqrt{\frac{\delta_{m}}{\delta_{m} + 2 Z_c^\ast}} \frac{U_0}{U_e}\qquad\textrm{with}\,\,U_e = \sqrt{\frac{Y}{3\rho}}\,.
\end{equation}
Here, \(U_e\) is the velocity of transverse sound waves in the elastic medium with which the quantity \(M \equiv U_0/U_e\) can be interpreted as an elastic Mach number \footnote{Clearly, we could also have inverted the function \(f(\lambda_m^2) = 2\lambda_m^2 +\lambda_m^{-4} -3\) to obtain an expression for the full curve \(\lambda_m^2 = g[(\delta_{m}/(\delta_{m} + 2 Z_c^\ast)) U_0^2/U_e^2]\), where \(g(\,\,)\) is just the inverse of \(f(\,\,)\).}. 

In Fig.~\ref{fig:SpreadingDiameterVsMa}a we plot the dimensionless maximum spreading diameter \(d_m/D_0\) as a function of the modified elastic Mach number \((\delta_{m}/(\delta_{m} + 2 Z_c^\ast))^{1/2}(U_0/U_e)\) for an enlarged number of experiments including two granular substrates with average grain diameters of \(114\) and \(200\) \(\mu\)m respectively and packing fractions \(\varphi\) ranging from \(0.55\) to \(0.62\). Clearly, the data collapses reasonably well for smaller values of the modified elastic Mach number, but disperse when that value increases.

The reason is neglecting the effect of surface tension for the small values of the Young's modulus. To incorporate surface tension we again follow \cite{Arora2018} and add a surface energy term \(E_c\) to the balance, which in the large deformation limit may be approximated as the product of the surface tension coefficient $\gamma$ and the sum of the upper and lower surface areas of the pancake shaped deformed sphere, i.e, \(E_c \approx \tfrac{1}{2}\pi D_0^2\gamma\lambda_m^2\). Now the sum of the elastic and capillary energies need to be equated to the portion of the impact kinetic energy available for the deformation of the sphere: \(E_d \approx E_e + E_c\) or
\begin{equation}\label{eq:enbal}
\frac{\pi D_0^3}{12}\frac{\delta_{m}}{\delta_{m} + 2 Z_c^\ast} \rho U_0^2 \approx  \frac{\pi D_0^3}{6}\left(G + \frac{3\gamma}{D_0}\right)\lambda_m^2\,,
\end{equation}
where the term in parenthesis may be written as \((G + 3\gamma/D_0) = 3Y(1+l_{ec}/D_0)\), where \(l_{ec} \equiv 9\gamma/Y\) is the elastocapillary length. Clearly, the surface tension term becomes important when \(l_{ec} \gtrapprox 1\). Solving the energy balance now leads to
\begin{equation}\label{eq:lambdam2}
\lambda_m \approx \frac{1}{\sqrt{2}}\sqrt{\frac{\delta_{m}}{\delta_{m} + 2 Z_c^\ast}} \frac{U_0}{U_{ec}}\qquad\textrm{with}\,\,U_{ec}^2 = U_e^2 + U_c^2 = \frac{Y}{3\rho} + \frac{3\gamma}{\rho D_0}\,,
\end{equation}
where the quantity \(U_0/U_{ec} = U_0/\sqrt{U_e^2 + U_c^2}\) can now be interpreted as an elastocapillary Mach number. Using the surface tension of water \(\gamma = 7.3 \cdot 10^{-2}\) N/m, we plot \(\lambda_m = d_m/D_0\) as a function of the modified elastocapillary Mach number \((\delta_{m}/(\delta_{m} + 2 Z_c^\ast))^{1/2}(U_0/U_{ec})\) in Fig.~\ref{fig:SpreadingDiameterVsMa}b and observe that the data now beautifully collapse onto a single curve. This implies that the spreading of the hydrogel spheres impacting onto a granular bed can be described as that of hydrogel spheres impacting on solid substrates, provided that one takes into account that only a portion \(E_d\) of the impact kinetic energy is available for droplet deformation.
   
 \begin{figure}
  \centerline{\includegraphics[width=\textwidth]{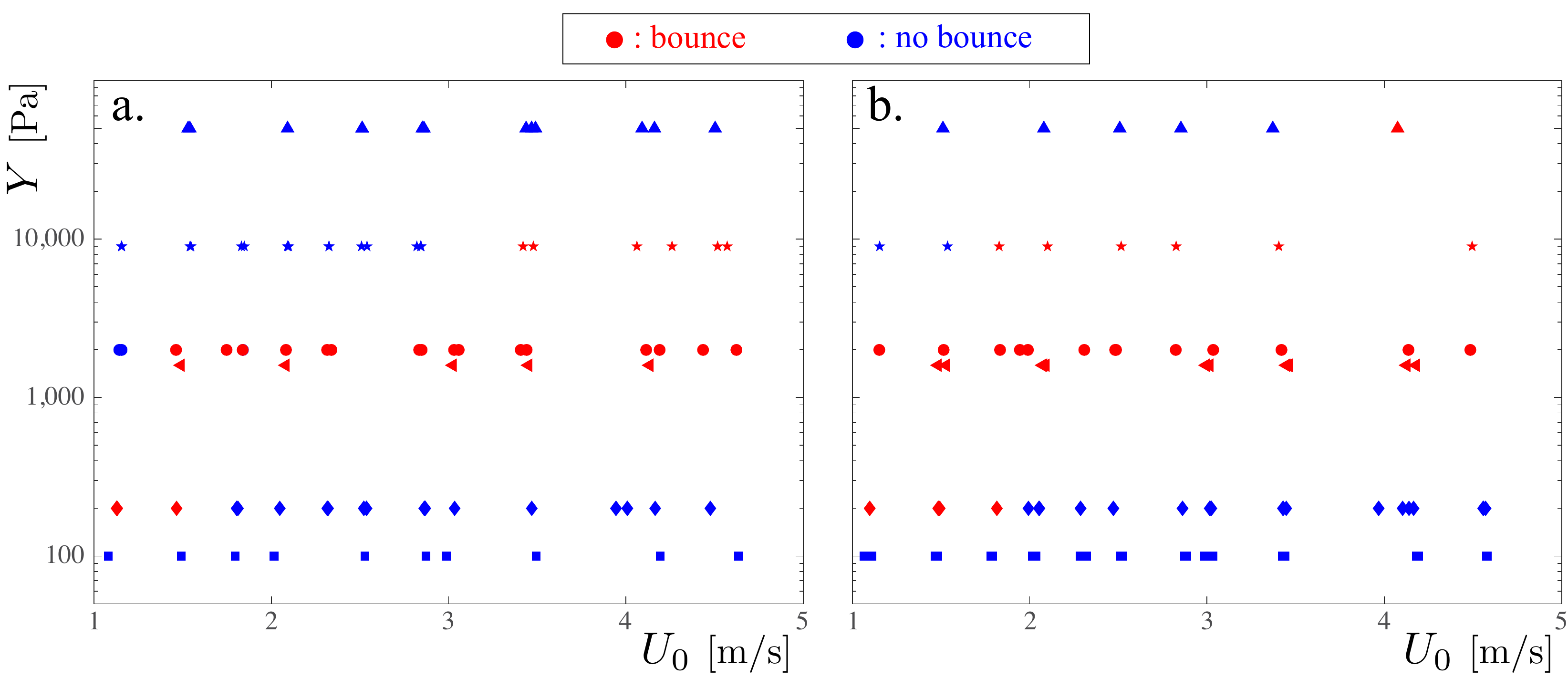}}
  \caption{Phase diagrams for the bouncing of hydrogel spheres after impact on a granular bed, presented in the parameter space created by the Young's modulus \(Y\) on the vertical and the impact velocity \(U_0\) on the horizontal axis. Red symbols indicate a rebound, whereas blue symbols represent impacts where bouncing is absent. The difference between the left and right plot is in the packing fraction of the granular substrate, which in (a) ranges from \(0.55\) to \(0.585\) and in (b) from \(0.585\) to \(0.62\). In both cases, the average grain diameter is \(200\) \(\mu\)m.}
 \label{fig:bouncingphasediagram}
  \end{figure}

\section{Bouncing}
\label{sec:bouncing}

The last subject we want to discuss is the post-impact motion of the hydrogel sphere. Clearly, an elastic sphere impacting on a solid substrate is likely to rebound~\citep{Tanaka2003} and also a droplet that impacts onto a plate may under certain circumstances rebound, depending on the wetting properties of the substrate~\citep{Josserand2016}. The situation becomes more complex if the substrate is a granular bed, i.e., if also the substrate is deformable, as it is in the current study.

From the snapshot series in Fig.~\ref{fig:snapshots}, it is clear that in series b a rebound occurs, whereas in a and c there is no evidence of bouncing. To better quantify the post-impact motion, we trace the motion of the hydrogel sphere by comparing the snapshot where the sphere first recovered its original shape after impacting and one frame later, from which one can obtain the vertical center-of-gravity positions \(h_{1}\) and \(h_{2}\). If \(h_{2}-h_{1}>0\), the sphere is observed to rebound. In Fig.~\ref{fig:bouncingphasediagram} we present two phase diagrams in the parameter space created by the sphere's Young's modulus \(Y\) and its impact velocity \(U_0\), for two different packing fraction ranges \(0.55 < \phi < 0.585\) (Fig.~\ref{fig:bouncingphasediagram}a) and  \(0.585 < \phi < 0.62\) (Fig.~\ref{fig:bouncingphasediagram}b). For both diagrams it is clear that bouncing only occurs at intermediate Young's moduli. Hydrogel spheres with large \(Y\) are prone to rebound at high impact energy, while spheres with small Young's modulus rebound at low impact energies. With the increase of packing fraction, the region where bouncing occurs is observed to expand towards larger Young's modulus and smaller impact energies. Physically, this implies that a stiff sphere will rebound when its deformation is large enough (i.e., at high impact energies), which in turn is influenced by the deformability of the granular bed: if the bed is relatively compact (\(0.585 < \phi < 0.62\)) a certain impact velocity may be sufficient  to trigger a rebound, whereas for a loose bed (\(0.55 < \phi < 0.585\)) an even larger impact velocity is needed, because the large deformability of the bed limits the deformation of the sphere.

\begin{figure}
  \centerline{\includegraphics[width=0.7\textwidth]{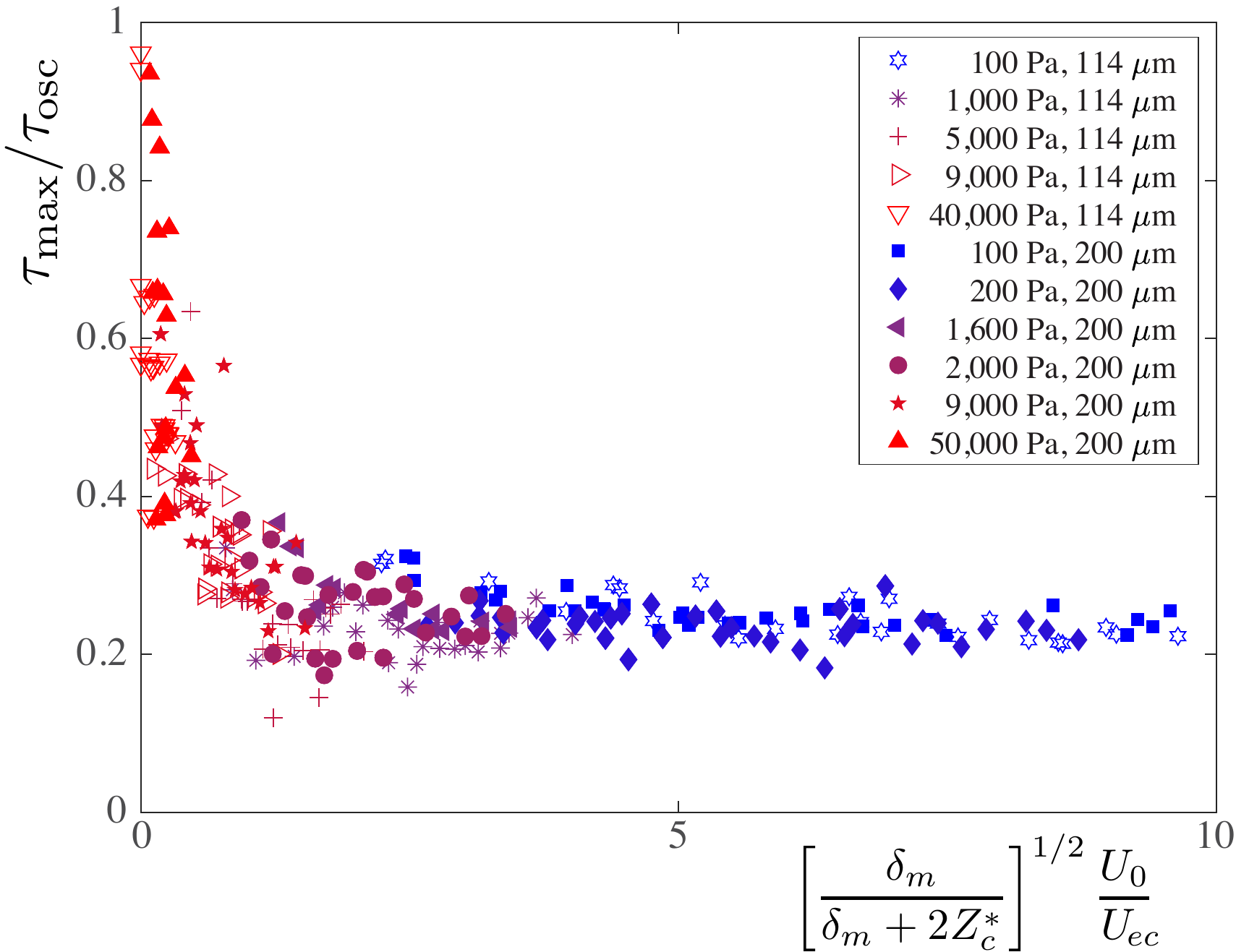}}
  \caption{Measured spreading time \(\tau_\textrm{max}\) scaled by one quarter of the theoretical oscillation time (\(\tau_\textrm{osc}\)) plotted versus the modified elastocapillary Mach number \((\delta_{m}/(\delta_{m} + 2Z_{C}^\ast))^{1/2}(U_0/U_{ec})\). Note that the extended data set from Fig.~\ref{fig:SpreadingDiameterVsMa} has been used to create this plot, using the same symbols and color coding.  
 }
 \label{fig:taumaxovertauosc}
  \end{figure}

The above observations indicate that one of the important parameters that determine whether bouncing occurs or not, may well be the ratio of the spreading time and the duration of the impact. To quantify this ratio, we measure the spreading time \(\tau_\textrm{max}\) (defined as the time difference between the moment of impact and that at which the maximal deformation is reached) which is plotted in Fig.~\ref{fig:SpreadingDiameterVsEk}b for the same dataset as used in Fig.~\ref{fig:SpreadingDiameterVsEk}a. Clearly, the spreading time decreases significantly with increasing Young's modulus and only mildly depends on the impact energy, where \(\tau_\textrm{max}\) changes from about \(3.8\) ms for \(Y = 100\) Pa, through \(1.3\) ms for \(Y=2,000\) Pa, to \(0.7\) ms for \(Y=9,000\) Pa. Subsequently, we compare the measured spreading time \(\tau_\textrm{max}\) with the expected oscillation time \(\tau_\textrm{osc}\) for a hydrogel sphere with Young's modulus \(Y\) and surface tension \(\gamma\), namely 
\begin{equation}\label{eq:tauosc}
\tau_\textrm{osc} = \frac{\pi D_0}{2\sqrt{2}U_{ec}} = \frac{\pi D_0}{2\sqrt{2}\sqrt{ \frac{Y}{3\rho} + \frac{3\gamma}{\rho D_0}}}
\end{equation}
which is equal to one quarter of the oscillation period of the hydrogel sphere. In Fig.~\ref{fig:taumaxovertauosc} we plot the ratio \(\tau_\textrm{max}/\tau_\textrm{osc}\) as a function of the modified elastocapillary Mach number \((\delta_{m}/(\delta_{m} + 2Z_{C}^\ast))^{1/2}(U_0/U_{ec})\). 
Clearly, all data again collapse onto a single master curve and \(\tau_\textrm{max}/\tau_\textrm{osc}\) rapidly decreases towards a constant value for large Mach number. 
This result is consistent with the experimental observations for a (rigid) plate impacted by droplets ~\citep{Chantelot2018}, and elastic balls ~\citep{Tanaka2003}. For the latter, the interpretation is that for globally deformed spheres the ratio \(\tau_\textrm{max}/\tau_\textrm{osc}\) tends to a constant value whereas for decreasing Mach number we enter the regime where the deformation is local (the Hertzian regime), which leads to an increase of the ratio \(\tau_\textrm{max}/\tau_\textrm{osc}\). Note that the poor collapse observed in \cite{Tanaka2005} for larger values of the Mach number could be avoided in our case due to the incorporation of the surface energy, i.e., by the introduction of the (modified) elastocapillary Mach number as in \cite{Arora2018}.  
 
\begin{figure}
  \centerline{\includegraphics[width=0.7\textwidth]{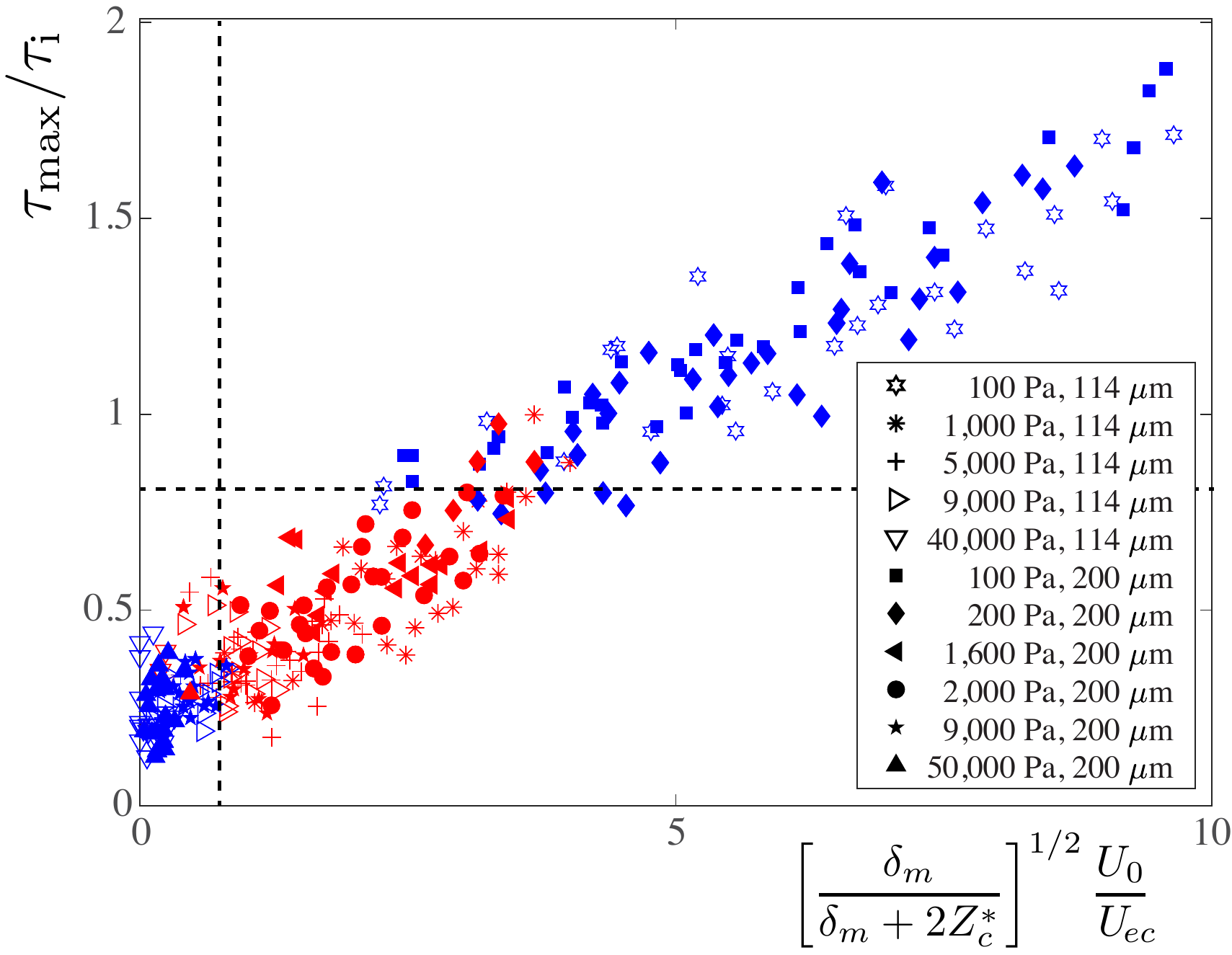}}
  \caption{Ratio of the spreading time \(\tau_\textrm{max}\) and the inertial time \(\tau_\textrm{i}=(\delta_{m}+2Z_{C}^\ast)/U_{0}\) versus the 
  modified elastocapillary Mach number \((\delta_{m}/(\delta_{m} + 2Z_{C}^\ast))^{1/2}(U_0/U_{ec})\). Note that the extended data set from Fig.~\ref{fig:SpreadingDiameterVsMa} has been used to create this plot, using the same symbols. This plot is a recast of the bouncing phase diagrams (Figs.~\ref{fig:bouncingphasediagram}) (for an extended dataset), where red symbols indicate a rebound and blue symbols represent impacts where bouncing is absent. The horizontal dashed black line located at \(\tau_\textrm{max}/\tau_\textrm{i} \approx 0.8\) provides an upper bound and the vertical dashed black line located at \((\delta_{m}/(\delta_{m} + 2Z_{C}^\ast))^{1/2}(U_0/U_{ec}) \approx 0.75\) a lower bound for the region in parameter space where bouncing occurs.
}
 \label{fig:timescaleratio}
\end{figure}

The inertial time scale \(\tau_\textrm{i}\), which for impact of an elastic sphere on a solid substrate can be estimated as \(\delta_{m}/U_{0}\), needs to be corrected for the deformation of the granular target as \(\tau_{i}=(\delta_{m}+2Z_{C}^\ast)/U_{0}\). With the above, we now are able to compute the ratio of the measured spreading time and the inertial time scale, \(\tau_\textrm{max}/\tau_\textrm{i}\), which allows us to recast the bouncing phase diagram of Figs.~\ref{fig:bouncingphasediagram}a and b, together with additional data for the granular bed consisting of smaller grains, into the new form presented in Fig.~\ref{fig:timescaleratio}.

First of all, all data appears to lie close to a straight line. This should be no huge surprise since both axes represent the ratio of an elastocapillary time scale and an inertial one. This is obvious for the vertical axis, but for the horizontal one may be realised by noting that the modified impact velocity is divided by the elastocapillary velocity scale, which combined with the fact that the length scale is determined by the size of the hydrogel sphere is equivalent to the ratio of an elelastocapillary and an inertial time scale. These are, however, obtained from very different and independent measurements.

Secondly, it is good to realise that the large time scale and velocity scale ratios in the upper right corner occur for large impact speeds and small Young moduli. So the softest hydrogel sphere impacts are located in the upper right quadrant and the stiffest ones close to zero, in the lower left quadrant of the plot.

The third observation to be made from Fig.~\ref{fig:timescaleratio} is that bouncing only occurs at intermediate values of the time scale ratio. We will now discuss why this happens. The upper bound, indicated by the horizontal dashed line (located at  \(\tau_\textrm{max}/\tau_\textrm{i} \approx 0.8\)), indicates the point at which the time scale ratio is of order unity, that is, where the spreading time \(\tau_\textrm{max}\) is equal to the inertial time \(\tau_\textrm{i}\). 

The reason is the following: Already \cite{richard2000} noted for the impact of a droplet on a solid substrate that rebound may not occur either due to dissipative mechanisms, such as viscous dissipation and contact angle hysteresis, or due to the fact that the vertical kinetic impact energy is converted partly into pure oscillation of the droplet and unavailable for rebound. Clearly, if the impacting object's spreading time is larger than the impact time, then all impact energy is necessarily converted into (horizontal) oscillation energy and no rebound will occur, which explains the location of the upper bound. This boundary should also be observable for hydrogel spheres bouncing from a solid substrate. For the case of hydrogel sphere impact on a granular substrate we have identified an additional mechanism that will act to decrease the location of the upper bound, namely that, especially for soft spheres, a layer of grains attaches to the surface of the sphere, making its mass increase and therefore bouncing less likely.

The lower bound however is unique to impact on granular targets, and appears because of the deformability of the substrate. When the elastic energy stored in the sphere releases during the penetration of the sphere inside the sand, due to the downwards motion of the granular substrate that energy cannot be converted into a vertical motion. The granular material is simply too soft for the sphere to rebound on. This happens when the elastocapillary deformation speed of the sphere becomes larger than the impact velocity, which dictates the velocity with which the granular substrate starts to yield. Indeed, the lower bound appears to be best described by the criterium that the modified elastocapillary Mach number is of order unity, indicated by the vertical dashed line located at \((\delta_{m}/(\delta_{m} + 2Z_{C}^\ast))^{1/2}(U_0/U_{ec}) \approx 0.75\).

For this case there exists an interesting parallel with the drop trampoline of \cite{Chantelot2018}, in which a simplified model is presented to describe the bouncing of a droplet on an elastic membrane in which both are described as a mass-spring combination. In their model, when (for comparable masses) the spring constant of the substrate is taken sufficiently smaller than the spring constant of the impactor, the velocity of the latter after the first half oscillation will never be larger than zero. Recalling that the deformation speed is proportional to the square root of the spring constant, this implies that the impactor will not rebound if the deformation speed of the substrate is sufficiently smaller than that of the impactor.

\section{Conclusion}
\label{sec:conclusion}

We systematically studied the impact of hydrogel spheres with Young's modulus varying over almost three orders of magnitude onto a granular bed. We observed how the impact speed and Young's modulus influence the diameter and depth of the crater that is created during impact, and compared the results with those obtained in our earlier work \citep{Jong2020} for the impact of liquid droplets and solid, undeformable spheres. To obtain a reliable comparison, we argued it to be essential to determine the portion of the impact kinetic energy that was invested in the formation of the crater by devising a method analogous to the one used for the impact of liquid drops \citep{Zhao2015}. 

Regarding the  crater diameter, the hydrogel sphere data lie between the limits set by solid and liquid impactors: For similar modified impact energies \(E_s\), large Young moduli spheres create narrow craters, slightly wider but comparable to those of solid spheres, whereas spheres with a small Young's modulus create wider craters that remain slightly narrower than those observed for the liquid impactors. The crater diameter data are consistent with a power law in the impact energy with exponents gradually varying from \(0.17\) to \(0.21\) from small to large Young moduli, consistent with the values found for liquid droplets (\(0.17-0.18\)) to solid spheres (\(0.20-0.24\)), where we note that the power exponent drops more sharply when \(Y\) becomes of the order of $200$ Pa and smaller. 

Regarding the crater depth the ordering of the data is as expected from the behavior of solid and liquid impactors, where small Young's modulus spheres create shallower impact craters than those with large Young moduli. In general however, impact craters resulting from hydrogel sphere impact are markedly shallower even than those created by liquid droplets, and even those created by the largest Young's modulus (\(Y=50,000\) Pa) are up to a factor 2 shallower than those created by the solid impactor. This could possibly be connected to the fact that the sand grains are observed to easily stick to the surface of the hydrogel particles which may cause an increased friction experienced by the hydrogel spheres during penetration into the sand bed, leading to a smaller final depth.

Turning to the deformation of the hydrogel spheres we show that, for a very broad range of impact parameters including two granular substrates consisting of differently sized grains and a full range of packing fractions, our data can be collapsed onto a single curve by the introduction of an elastocapillary Mach number which rescales the impact velocity with the typical elastocapillary speed introduced in \cite{Arora2018}, provided that we correct the impact velocity for the fact that only part of the impact kinetic energy is used for the deformation of the spheres. 

Finally, we observe that hydrogel spheres may bounce off the granular substrate only for a certain intermediate range of Young moduli and impact speeds. We trace this behavior back to the ratio of the spreading time and the impact time: For large values of this ratio a rebound is not expected to occur because most of the energy will be stored in the horizontal direction, since most of the spreading is occurring while the vertical deformations have already reached their maximum. This threshold, happening when the time ratio is of order unity, would exist for both a solid and a deformable substrate. For very small time-ratio values a rebound is not expected because the granular substrate is still in the process of moving down while the elastic sphere is already relaxing back to its original spherical shape, such that the elastic energy will not be able to propel the sphere into the upward direction. This happens uniquely for a deformable substrate, and can be traced back to the condition where the modified elastocapillary Mach number is of order unity.

\section*{Acknowledgements}
The authors want to thank GertWim Bruggert, Diana Garc{\'i}a-Gonz{\'a}lez, Kirsten Harth, Rianne de Jong, Song-Chuan Zhao, Ty Phou, Lingna Yang, Pierre Chantelot and Rana Depti for their invaluable help in rebuilding the setup, producing hydrogel particles and valuable discussions. XY acknowledges financial support National Natural Science Foundation of China (No.11872029).

\bibliographystyle{jfm}
\bibliography{hydrogel_impact}

\end{document}